Strongly Cavity-Enhanced Spontaneous Emission from Silicon-Vacancy Centers in Diamond

*Jingyuan Linda Zhang[1,‡], Shuo Sun[1,‡], Michael J. Burek[2,‡], Constantin Dory[1], Yan-Kai Tzeng[5], Kevin A. Fischer[1], Yousif Kelaita[1], Konstantinos G. Lagoudakis[1], Marina Radulaski[1], Zhi-Xun Shen[3,4,5], Nicholas A. Melosh[3,4], Steven Chu[5,6], Marko Lončar[2], Jelena Vučković[1]*

[1]*E. L. Ginzton Laboratory, Stanford University, Stanford, California 94305, USA*

[2]*School of Engineering and Applied Sciences, Harvard University, Cambridge, Massachusetts 02138, USA*

[3]*Geballe Laboratory for Advanced Materials, Stanford University, Stanford, California 94305, United States*

[4]*Stanford Institute for Materials and Energy Sciences, SLAC National Accelerator Laboratory, Menlo Park, California 94025, USA*

[5]*Department of Physics, Stanford University, Stanford, California 94305, USA*

[6]*Department of Molecular and Cellular Physiology, Stanford University, Stanford, California 94305, USA*

ABSTRACT

Quantum emitters are an integral component for a broad range of quantum technologies including quantum communication, quantum repeaters, and linear optical quantum computation. Solid-state color centers are promising candidates for scalable quantum optics due to their long coherence time and small inhomogeneous broadening. However, once excited, color centers often decay through phonon-assisted processes, limiting the efficiency of single photon generation and photon mediated entanglement generation. Herein, we demonstrate strong enhancement of spontaneous emission rate of a single silicon-vacancy center in diamond embedded within a monolithic optical cavity, reaching a regime where the excited state lifetime is dominated by spontaneous emission into the cavity mode. We observe 10-fold lifetime reduction and 42-fold enhancement in emission intensity when the cavity is tuned into resonance with the optical transition of a single silicon-vacancy center, corresponding to 90% of the excited state energy decay occurring through spontaneous emission into the cavity mode. We also demonstrate the largest to date coupling strength ($g/2\pi = 4.9 \pm 0.3\,\text{GHz}$) and cooperativity ($C = 1.4$) for color-center-based cavity quantum electrodynamics systems, bringing the system closer to the strong coupling regime.

Solid-state quantum emitters such as color centers in solids[1] are suitable for implementing an on-chip integrated platform for many applications in quantum information processing,[2-4] including boson sampling,[5-7] quantum key distributions,[8, 9] as well as photonic interfaces for entanglement distribution.[10-13] Many color centers exhibit a spin degree of freedom with long coherence time,[14-18] which can be used as optically addressable spin qubits. Compared to other widely studied solid-state quantum emitters such as semiconductor quantum dots, they are particularly promising for scalable operations due to their small inhomogeneous broadening.[19]

In order to take advantage of their long spin coherence time and narrow inhomogeneous broadening, a cavity-based spin-photon interface is required to enhance the coherent emission of photons into the zero-phonon line (ZPL), which would improve the heralded entanglement generation rate. In addition, quantum emitters with their emission enhanced by cavities act as ultrafast single photon sources, which may find applications in high-repetition-rate quantum key distribution. However, in prior works which demonstrated enhancement of color center emission into the ZPL via resonant coupling with nanophotonic cavities,[20-29] the measured lifetime reduction has been limited due to the poor quantum efficiency of the emitter, small branching ratio into the ZPL, limited optical quality factor of the cavity, or low coupling strength between the emitter and the cavity.

In this work, we demonstrate 10-fold lifetime reduction combined with 42-fold intensity enhancement for individual color centers in diamond coupled to monolithic optical cavities, reaching a regime where spontaneous emission through the ZPL into the cavity mode dominates all other decay channels. Moreover, we demonstrate the largest to date coupling strength and cooperativity for color-center-based cavity quantum electrodynamics systems. We use negatively charged silicon-vacancy (SiV⁻) color centers in diamond, grown by chemical vapor deposition (CVD),[30, 31] embedded within nanofabricated photonic crystal cavities. The resulting SiV⁻ centers do not exhibit significant spectral diffusion, with linewidths comparable to those reported in bulk diamond and in nanobeams.[19, 32] A high yield of emitter-cavity systems displaying strong enhancement is observed, based on measurements of cavities nearly resonant with the ZPL emission.

Fabrication of emitter-cavity systems began with a single-crystal diamond plate (Type IIa, < 1 ppm [N], Element Six), on which a nominally 100-nm-thick layer of diamond containing SiV⁻ centers was grown homoepitaxially via microwave plasma chemical vapor deposition (MPCVD).[33] Silicon atoms are readily available in the growth chamber during this MPCVD step, due to hydrogen plasma etching of a silicon carrier wafer placed underneath the diamond substrate. SiV$^-$ centers were subsequently formed *in situ* by silicon incorporation into the evolving diamond layer via plasma diffusion.

Nanophotonic cavities were fabricated in this silicon rich diamond using electron beam lithography (EBL) followed by angled-etching[34-36] of the bulk single-crystal diamond, with details given elsewhere[27] and presented in the Supplementary Information. Figures 1 (a) and (b) display a zoomed-in top and angled view of a typical fabricated nanophotonic cavity, respectively. The optical cavity architecture used in this work is a "nanobeam" photonic crystal cavity[37], formed by a one dimensional lattice of elliptical air holes along the freestanding waveguide. An optical cavity mode is localized in the structure by positively tapering the air hole major radius (perpendicular to the waveguide axis) from each end towards the center device mirror plane. Device dimensions (details in the Supplementary Information) were chosen to target a cavity mode near the ZPL emission of SiV$^-$ center in diamond at λ ~ 737 nm. The figures of merit for our nanobeam cavity design (obtained by simulation via finite-difference-time-domain (FDTD) methods) yield a theoretical quality factor of Q ~ 10,000, and wavelength scale mode volume of $V = 1.8\left(\frac{\lambda}{n}\right)^3$, where $n = 2.402$ is the refractive index of diamond. The cavity mode profiles are shown in Figure 1(c).

Optical characterization of fabricated devices was performed in a home built confocal microscope setup at cryogenic temperatures (~ 5 K). A low temperature photoluminescence (PL) spectrum from a representative device under 720 nm laser excitation is shown in Figure 1(d). Upon cooling to liquid helium temperature, four characteristic optical transitions between spin-orbit eigenstates (labeled A to D in Figure 1(d)) of the SiV$^-$ center are revealed. In the PL spectrum, the cavity mode is observed blue-detuned from the SiV$^-$ emission lines at ~ 734.5 nm, with a Q ~ 8300 extracted from the full width at half maximum (FWHM).

The SiV$^-$ optical transition C linewidth, characterized by photoluminescence excitation (PLE), approaches ~ 304 MHz (full-width-half-maximum) as the excitation power was reduced to minimize power broadening effects (Figure 2(a)). This is 3.6 times the Fourier-transform limited natural linewidth, and is comparable to that of ion-implanted SiV$^-$ centers.[32] The linewidth is slightly broader than that of CVD grown SiV⁻s in bulk,[19] which could be due to nanofabrication induced strain and proximity to surfaces. We confirm the single photon nature of the emission through a second-order correlation measurement of the same SiV⁻ emission in our emitter-cavity system under non-resonant pulsed laser excitation from a Ti:sapphire laser, which yields a $g^{(2)}(0) = 0.04$ (Figure 2(b)).

Purcell enhancement of the SiV⁻ emission is observed as our cavity mode was tuned into resonance with the individual dipole transitions of the SiV⁻ center. Figure 3(a) shows the device PL spectra as we continuously red-shifted the cavity wavelength by gas tuning (for details see Supplementary Information). Observed emission intensities of individual SiV⁻ dipole transitions resonantly coupled to the optical cavity were strongly enhanced due to the Purcell effect. Figure 3(b) displays two spectra collected at the detuning conditions indicated by the colored dashed lines in Figure 3(a), where the optical cavity was far detuned from (green) and on resonance with (blue) transition B. With the cavity on resonance, transition B exhibits an emission intensity increase by a factor of ~ 42.4 compared to the far detuned case. We have not been able to saturate transition B under resonant condition with up to of 1.8 mW excitation power, limited by our ability to stabilize the cavity frequency under high excitation power. With a maximum excitation power of 1.8 mW, we collect $1.33 \times 10^5$ counts per second from transition B when it is resonant with the cavity.

To quantitatively explore this observed Purcell enhancement further, measurements of the SiV⁻ center spontaneous emission rate were performed with the cavity both on and off resonance. When the cavity was far detuned, with the temporal profile shown in Figure 3(c), the spontaneous decay rate is extracted from a single exponential fit to be $\tau_{off} = 1.84 \pm 0.04$ ns. When the cavity was tuned on resonance with transition B, time-resolved spectroscopy was performed with a streak camera (Hamamatsu C5680), which has a faster instrument response time (<5 ps) compared to that of hundreds of picoseconds for a single photon counting module (SPCM). The intensity of the cavity enhanced transition B dominates all other emission lines (blue curve, Figure 3(b)), such that we are able to select the corresponding spectral region on the streak camera image, as shown by the dotted box in Figure 3(e). Fitting this binned luminescence to a single exponential decay yields a significantly decreased resonant lifetime of $\tau_{on} = 194 \pm 8$ ps (Figure 3(d)). For this device, we focus on the spontaneous emission lifetime when the cavity is resonant with transition B, because this transition exhibits the largest improvement in brightness when resonant with the cavity. The 10-fold lifetime reduction combined with a 42-fold intensity increase on resonance implies a large Purcell factor. Because of the non-unity off-resonance branching ratio into transition B,[21-23, 25, 28, 38-47] the actual Purcell factor is even higher than the directly measured lifetime reduction $\tau_{off}/\tau_{on}$. Through quasi-resonant pumping and detection, an upper bound for the off-resonance branching ratio of 0.325 was measured, corresponding to a minimum Purcell factor of $F_{min} = 26.1 \pm 1.8$[23] (details of both the branching ratio measurement and the Purcell factor calculation are in the Supplementary Information).

Furthermore, we demonstrate that the strong Purcell enhancement leads to a regime where spontaneous emission through the ZPL into the cavity mode dominates all other decay channels. We use the $\beta$-factor to characterize the fraction of the excited state energy decay through spontaneous emission into the cavity mode, defined as $\beta = 1 - \tau_{on}/\tau_{off}$. The $\beta$-factor scales from 0 to 1, with 1 being the excited state lifetime completely determined by the spontaneous emission into the cavity mode, and 0 being the excited state not emitting into the cavity at all. We estimate the $\beta$-factor to be 89.7±0.6%, demonstrating that the lifetime of the excited state is now dominated by the spontaneous emission into the cavity enhanced zero-phonon line. The large $\beta$-factor combined with the short lifetime of 194 ps, yields a single photon emission rate $\beta/\tau_{on}$ of ~ $2\pi \cdot 0.74$ GHz into the cavity mode. Therefore, our system shows potential for an ultrafast, nearly gigahertz single photon source.

We also show that the strong light-matter coupling enables a coherent dipole induced transparency effect.[48] We use a tunable continuous-wave laser to excite the device from the notch at one end of the waveguide, and collect the transmitted light from an identical notch at the other end. Figure 4(a) shows the transmission spectrum of the bare cavity mode, where the cavity is far detuned from all four transitions of the SiV⁻. The bare cavity shows a single Lorentzian lineshape that strongly suppresses transmission at the cavity resonance. In contrast, when we tune the cavity into resonance with transition B of the SiV⁻, we observe a clear transmission peak at transition B due to dipole induced transparency. By fitting the measured data (blue dots) to a numerical model (red solid line; fitting see Supplementary Information), we extract the coupling strength between transition B and the cavity to be $g/2\pi = 4.9 \pm 0.3$ GHz, and the cavity energy decay rate $\kappa/2\pi = 49.7 \pm 2.0$ GHz. We also calculate the cooperativity of the system, defined as $C = 4g^2/\kappa\gamma$

, to be $C = 1.4$, where $\gamma/2\pi = 1.36 \pm 0.06\,\text{GHz}$ is the linewidth of transition B when the cavity is far detuned, obtained through photoluminescence excitation measurements.[49] To the best of our knowledge, both the coupling strength and the cooperativity are the highest values reported so far for color center based cavity quantum electrodynamics systems. However, the origin of the superior performance of our devices compared to the previous works[26] requires further investigation. We also determine the coupling strength between transition C and the cavity to be $1.4 \pm 0.1\,\text{GHz}$ by performing the same measurements on transition C (see Supplementary Information). We attribute the difference in the coupling strength for transitions B and C to strain induced difference in the selection rules.[50] In fact, this particular device we measured shows a much larger ground state splitting ($155.0 \pm 2.5\,\text{GHz}$) compared with typical values, suggesting a large strain is present at the emitter location.[18]

Finally, we observe a high yield of the strongly cavity enhanced SiV⁻ centers, as summarized in Table 1. Eight devices were found displaying cavity resonances blue detuned by several nanometers from the SiV⁻ ZPL emission, which is within the tuning range of the gas tuning method. Of the eight devices, four contain stable SiV⁻ centers with low spectral diffusion, with the cavity tuning range reaching the ZPL wavelengths. In Table 1, $\tau_{on}$, $\tau_{off}$, $\tau_{off}/\tau_{on}$, $I_{on}/I_{off}$, and $\beta$ are the on-resonance lifetime, off-resonance lifetime, lifetime reduction factor, intensity increase factor, and fraction of the excitation decay through the spontaneous emission into the cavity mode, respectively. These four systems all exhibit a lifetime reduction greater than 5.5, and a $\beta$-factors greater than 82%.

In summary, we have demonstrated Purcell enhancement of single photon emission from as-grown SiV⁻ centers in diamond by coupling them to monolithic photonic crystal cavities. The cavity coupled SiV⁻ centers exhibit ~10-fold lifetime reduction, from which we extract a $\beta$-factor of 89.7±0.6%, an emitter-cavity coupling strength of $g/2\pi = 4.9 \pm 0.3\,\text{GHz}$, and a cooperativity of 1.4. All the parameters represent state-of-the-art values for color-center-based cavity quantum electrodynamics systems. The large $\beta$-factor suggests promising potential for scalable single photon sources operating at the gigahertz regime. Further work on improving the extraction efficiency of the coupled system, through either far-field optimization for free space extraction[51] or through an efficient fiber-coupled diamond nanophotonic interface,[27] would push SiV⁻ closer towards scalable quantum networks. This platform could also be readily extended to other color centers such as the germanium-vacancy centers[52, 53] and neutral silicon-vacancy center in diamond,[54] which also exhibit desirable optical properties and hold great promise for quantum information processing. The high Purcell enhancement and large coupling strength demonstrated in our system brings us closer to reaching the strong coupling regime, by either improving the cavity parameters[55] (i.e. improving Q by a factor of 2 and decreasing V by a factor of 1.5) or by incorporating multiple (~7) emitters.[56-58]

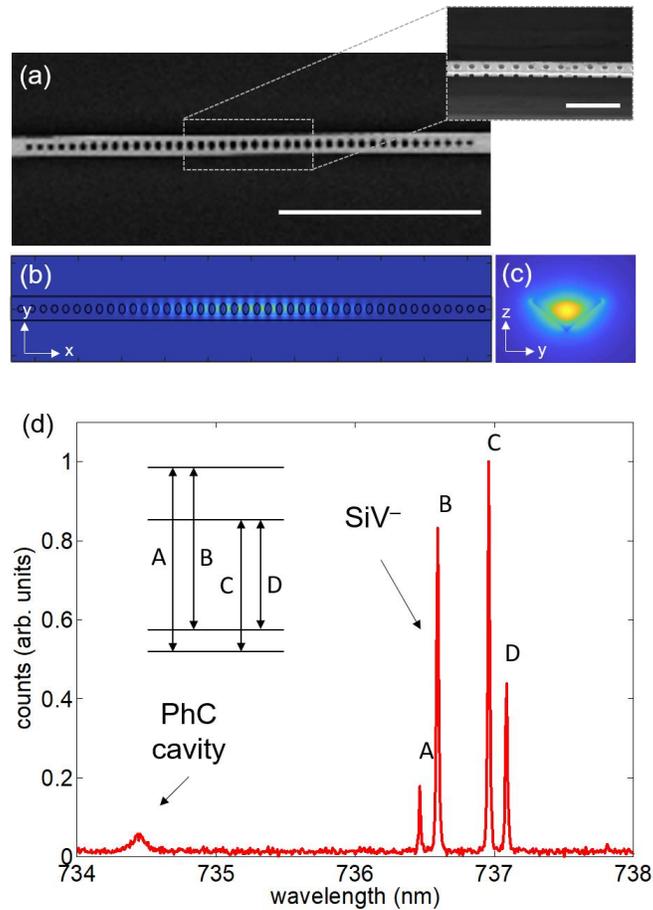

**Figure 1. High-Q nanobeam photonic crystal cavity.** (a) Scanning electron microscopy (SEM) images of a nanobeam photonic crystal (PhC) cavity fabricated from single crystal diamond, with the inset showing the angled-view of the cavity region. Scale bars in (a) and the inset: 5 $\mu$m and 1 $\mu$m respectively. (b) Electric field intensity profile of the fundamental cavity mode of the photonic crystal cavity. (c) Cross-sectional electric field intensity profile of the fundamental cavity mode of the photonic crystal cavity, taken at the center plane in the x-direction. (d) Low temperature photoluminescence (PL) spectrum of a SiV$^-$ center and the cavity mode. The four narrow lines correspond to the four optical transitions of a SiV$^-$, as shown by the double arrows in the level structure in the inset. The cavity mode is blue-detuned from the SiV emission at ~ 734.5 nm.

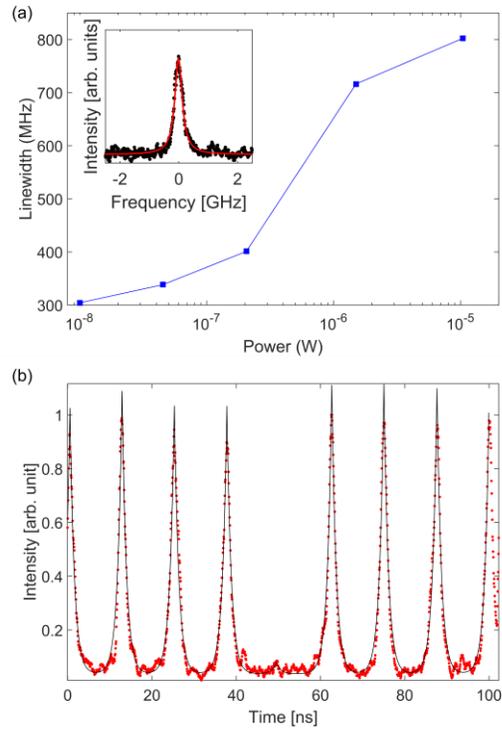

**Figure 2. Emission properties of the single SiV⁻ centers.** (a) The linewidth of transition C of a SiV⁻ in the nanobeam photonic crystal cavity. The linewidth at low excitation power reaches 304 MHz, as shown in the inset. (b) Second-order autocorrelation measurement of the cavity coupled SiV⁻ center emission under pulsed excitation, yielding $g^{(2)}(0) = 0.04$.

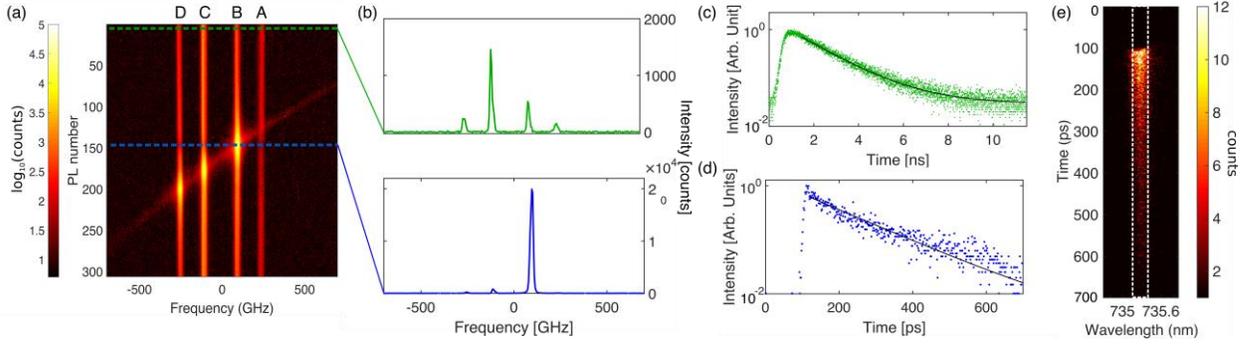

**Figure 3. Enhanced photoluminescence due to coupling to the photonic crystal cavity.** (a) High resolution PL spectra over the SiV⁻ emission region, as the cavity is tuned across the SiV⁻ emission through argon gas condensation. The resonant and detuned cases are taken at the blue and green dashed lines respectively. (b) High resolution PL spectra of the SiV⁻ center when the cavity is detuned from (green) and resonant with (blue) transition B of the SiV⁻. (c-d) Time-resolved photoluminescence measurements of transition B of the SiV⁻ yields a detuned lifetime $\tau_{off} = 1.84 \pm 0.04$ ns (c), and resonant lifetime $\tau_{on} = 194 \pm 8$ ps (d). (e) Time-resolved spectroscopy measurement of transition B on-resonance. In this streak camera image, the wavelength is dispersed in the horizontal direction by a grating and time is resolved in the vertical direction. The binned region is boxed by the dotted lines.

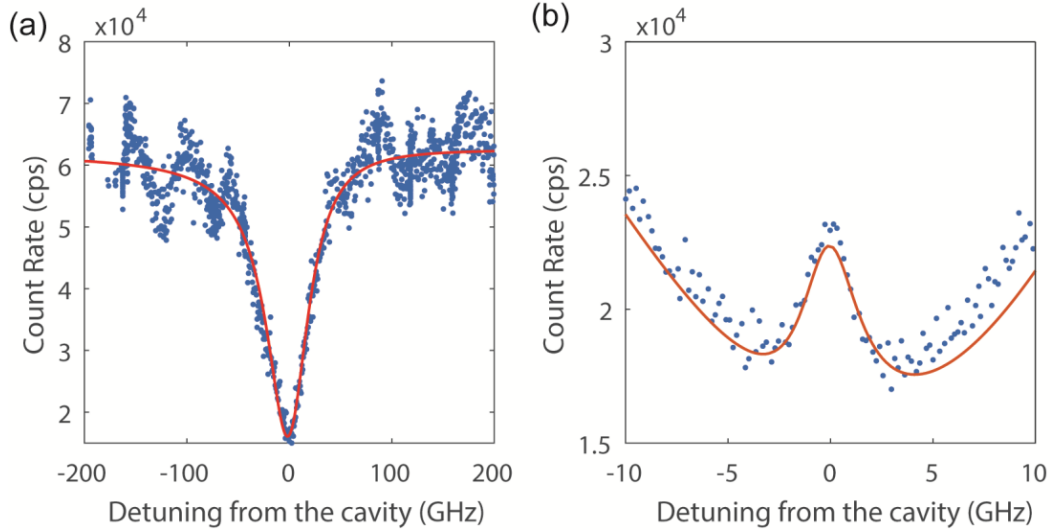

**Figure 4.** (a) Cavity transmission spectrum when the cavity is far detuned from all transitions of the SiV⁻. (b) Dipole induced transparency peak in the transmission spectrum when the same cavity is resonant with transition B of the SiV⁻. In both panels, blue dots are measured data, and red solid lines are fit to a numerical model.

**Table 1:** Purcell enhancement parameters of the SiV⁻ centers

| SiV⁻ # | $\tau_{on}$ [ns] | $\tau_{off}$ [ns] | $\tau_{off}/\tau_{on}$ | $I_{on}/I_{off}$ | $\beta$ (%) |
|---|---|---|---|---|---|
| 1 | 0.340±0.017 | 1.88±0.02 | 5.5±0.3 | 17.7 | 82.4±1.0 |
| 2 | 0.208±0.011 | 1.79±0.02 | 8.6±0.6 | 5.6 | 88.6±0.7 |
| 3 | 0.194±0.008 | 1.84±0.04 | 9.5±0.6 | 42.4 | 89.7±0.6 |
| 4 | 0.158±0.003 | 1.70±0.02 | 10.8±0.3 | 39.1 | 91.0±0.3 |

**Supporting Information**.

Material synthesis and device fabrication, Optical measurement set-up, Minimum Purcell factor $F_{\min}$, Measurement of $\xi_{\max}$, Comparison to the theoretical Purcell factor, Analysis of non-radiative decay and its impact on beta factor, Theoretical model for cavity transmission spectrum.

*Jingyuan Linda Zhang: ljzhang@stanford.edu

**Acknowledgement**
We thank Mikhail Lukin for fruitful discussions. This work is financially supported by the Department of Energy (DOE), Laboratory Directed Research and Development program at SLAC National Accelerator Laboratory, under contract DE-AC02-76SF00515, Army Research Office (ARO) (Award no. W911NF1310309), Air Force Office of Scientific Research (AFOSR) MURI Center for Quantum Metaphotonics and Metamaterials, ONR MURI on Quantum Optomechanics (Award No. N00014-15-1-2761), National Science Foundation (NSF) EFRI ACQUIRE program (Award No. 5710004174), and the Army Research Laboratory. This work was performed in part at the Stanford Nanofabrication Facility of National Nanotechnology Infrastructure Network (NNIN) supported by the National Science Foundation (ECS-9731293 and DMR-1503759), and at Stanford Nano Shared Facility. Device fabrication was performed in part at the Center for Nanoscale Systems (CNS), a member of the NNIN, which is supported by the National Science Foundation under NSF award no. ECS-0335765. CNS is part of Harvard University. CD acknowledges support from the Andreas Bechtolsheim Stanford Graduate Fellowship. KAF acknowledges support from the Lu Stanford Graduate Fellowship and the National Defense Science and Engineering Graduate Fellowship.